\documentclass[10pt]{article}
\usepackage{mhchem}
\usepackage{fancyhdr}
\usepackage{extramarks} 
\usepackage{amsmath}
\usepackage{amsthm}
\usepackage{amsfonts}
\usepackage{siunitx}
\usepackage{tikz}
\usepackage[plain]{algorithm}
\usepackage{algpseudocode}
\usepackage{multirow}
\usepackage{booktabs}
\usepackage{palatino}
\usepackage{graphicx}
\usepackage{subfigure}

\usepackage{blindtext}

\usepackage[colorlinks,linkcolor=black,anchorcolor=black,citecolor=black,urlcolor=blue]{hyperref}
\usepackage{amsmath,bm}
\usepackage{booktabs}
\usepackage{mathtools}
\usepackage{amssymb}
\usepackage{caption}
\usepackage{capt-of}
\usepackage{mciteplus}
\usepackage{cite}
\usepackage{mathrsfs}
\usepackage[title,titletoc,toc]{appendix}
\usepackage{xr}
\usepackage{parskip}
\usepackage{soul}
\usepackage{textcomp}
\usepackage[colaction]{multicol}
\usepackage[switch]{lineno}
\usepackage{lipsum}
\usepackage{etoolbox}
\usepackage{longtable}
\usepackage{array}
\usepackage{tablefootnote}
\usepackage{graphicx}
\usepackage{color, colortbl}
\usepackage{hhline}

\definecolor{Lightblue}{rgb}{0.867,0.914,0.961}
\definecolor{Lightgreen}{rgb}{0.883,0.934,0.848}

\newcolumntype{C}[1]{>{\centering\arraybackslash}p{#1}}
\usetikzlibrary{automata,positioning}
\usetikzlibrary{automata,positioning}

\begin{document}

\title{Machine learning analysis of cocaine addiction informed by DAT, SERT,
and NET-based interactome networks}
\author{Hongsong Feng$^1$, Kaifu Gao$^1$, Dong Chen$^1$,\\ Alfred J  Robison$^2$, Edmund Ellsworth$^3$ and Guo-Wei Wei$^{1,4,5}$\footnote{
 		Corresponding author.		Email: weig@msu.edu} \\
 $^1$ Department of Mathematics, \\
 Michigan State University, MI 48824, USA.\\
 $^2$ Department of Physiology,\\
Michigan State University, MI 48824, USA.\\
 $^3$Department of Pharmacology \& Toxicology\\
Michigan State University, MI 48824, USA.\\
$^4$Department of Electrical and Computer Engineering,\\
 Michigan State University, MI 48824, USA. \\
 $^5$ Department of Biochemistry and Molecular Biology,\\
 Michigan State University, MI 48824, USA. \\
 }

\date{\today} 

\maketitle

\begin{abstract}
Cocaine addiction is a psychosocial disorder induced by the chronic use of cocaine and causes a large of number deaths around the world. Despite many decades’ effort, no drugs have been approved by the Food and Drug Administration (FDA) for the treatment of cocaine dependence. Cocaine dependence is neuro- logical and involves many interacting proteins in the interactome. Among them, dopamine transporter (DAT), serotonin transporter (SERT), and norepinephrine transporter (NET) are three major targets. Each of these targets has a large protein-protein interaction (PPI) network which must be considered in the anti-cocaine addiction drug discovery. This work presents DAT, SERT, and NET interactome network-informed machine learning/deep learning (ML/DL) studies of cocaine addiction. We collect and analyze 61 protein targets out 460 proteins in the DAT, SERT, and NET PPI networks that have sufficient existing inhibitor datasets. Utilizing autoencoder and other ML algorithms, we build ML/DL models for these targets with 115,407 inhibitors to predict drug repurposing potentials and possible side effects. We further screen their absorption, distribution, metabolism, and excretion, and toxicity (ADMET) properties to search for nearly optimal leads for anti-cocaine addiction. Our approach sets up a systematic protocol for artificial intelligence (AI)-based anti-cocaine addiction lead discovery.

\end{abstract}


\pagenumbering{roman}
\begin{verbatim}
\end{verbatim}

%
\newpage

\setcounter{page}{1}
\renewcommand{\thepage}{{\arabic{page}}}

\section{Introduction}

Cocaine abuse is a serious health concern in the United States (US) and around the world. It is not  only associated with a series of medical complications including increased risk of HIV (human immun- odeficiency virus), hepatitis B, and heart disease, but also is associated with rising rates of crime and violence \cite{morris1991cocaine, chaisson1989cocaine,carvalho2009crack}. Despite significant attention to discover effective pharmacotherapies for the treatment of cocaine dependence, no effective medication has been approved by the US Food and Drug Administration (FDA).
 
 Cocaine is a tropane alkaloid and stimulant drug with significant addictive potential. It is a non- selective inhibitor of monoamine transporters including dopamine (DAT), serotonin (SERT), and norepinephrine (NET) transporters \cite{beuming2008binding,kristensen2011slc6,elliott2005psychostimulants}. By binding to these transporters, cocaine blocks reup-take of dopamine, serotonin, and norepinephrine, leading to higher synaptic and extracellular concentrations of these critical neurotransmitters. Cocaine elicits psychostimulant activities through increased activation of the monoamine receptors on post-synaptic neurons and can cause enhanced euphoric experiences.

The rewarding and addictive effects of psychostimulants are directly associated with the increased levels of dopamine in the nucleus accumbens (NAc) \cite{tzschentke2000functional}, which is a critical component of mesolimbic and mesocortical dopamine pathways. This pathway originates from the ventral tegmental area of the mid- brain and terminates with dopamine release in NAc \cite{koob2000neurobiology}, and makes significant contributions to stimulant reward \cite{tzschentke2000functional}. DAT is considered to play a primary role in the addictive effect of cocaine. Due to critical role of DAT in cocaine addiction, many experimental medications have been designed to target the dopamine system.

In addition to DAT, SERT also plays important roles in cocaine effects. \textit{In vivo} experiments in rats showed that enhanced dopamine transmission by acute cocaine intoxication in the nucleus accumbens is accompanied by elevated release of serotonin \cite{broderick2004clozapine}. Moreover, cocaine withdrawal is associated with de- creased serotonin in nucleus accumbens in microdialysis studies \cite{dworkin1995rat,parsons1995extracellular}. Mice with genetic deletion of DAT still show the rewarding effects of cocaine and cocaine conditioned place preference \cite{rocha1998cocaine}, which suggests non-DAT targets contribute to psychostimulant effects. However, combined dopamine and serotonin transporter knockouts eliminates cocaine place preference in mice, indicating that SERT makes a key contribution \cite{sora2001molecular,filip2005serotonergic}.

Serotonin neurons originate in the raphe nuclei of the mid-brain and are found in various regions of the brain including a dense innervation of terminals to ventral tegumental area (VTA) and nucleus accumbens (NAc)  \cite{halliday1989serotonin}. Cocaine-induced elevated extra-cellular levels of serotonin hyperactivate serotonin receptors in these and other brain regions. The actions of serotonin are mediated by at least 16 receptor subtypes that are grouped into seven families \cite{hoyer2002molecular}. Different receptors may serve different modulatory effects due to various neurochemical mechanisms with serotonin. Serotonin 1A receptor (5-HT$_{1A}$) is one of the most important receptors, existing pre- and post-synaptically in many rain areas \cite{vanderhoek1990evidence}, and is involved in nearly all serotonin mediated effects. Another family of serotonin receptors, including 5-$\rm HT_{2A}$ and 5-$\rm HT_{2C}$, is associated with impulsivity and cue reactivity to cocaine. Either selective 5-$\rm HT_{2A}$ antagonist or 5-$\rm HT_{2C}$ lessen impulsivity and cocaine-seeking in animal models, and the synergism of pharmacotherapeutics targeting these receptors was reported to attenuate a variety of aspects of cocaine relapse \cite{cunningham2013synergism}. Some experiments on animals showed that 5-HT$_{\text{3}}$ receptor antagonists also have potential therapeutic efficacy in curbing systems of cocaine consumption \cite{davidson20025,davidson2002ondansetron} or are effective in abolishing restatement of cocaine self-administration \cite{davidson2002ondansetron}. Preclinical studies in rats show that serotonin-enhancing medications could help decrease self-administration of cocaine \cite{carroll1990fluoxetine,peltier1993effects}, though selective serotonin reuptake inhibitors gave mixed results for treating cocaine addiction in clinical trials \cite{batki1996controlled}.

Many noradrenergic neurons are localized in brainstem nuclei, while noradrenergic axons project virtually 
everywhere in the brain \cite{smythies2005section}. NET is in the plasma membrane of noradrenergic neurons and plays a primary role in the inactivation of noradrenergic signaling by reuptake of synaptically released norepinephrine (NE). Cocaine causes elevated synaptic concentration of NE by its competitive binding to NET and subsequently increases activation of postsynaptic NE receptors \cite{tellioglu2001genetic,amara1998neurotransmitter}. NE is a crucial neurochemical messenger in central noradrenergic and peripheral sympathetic pathways, and its effects are mediated by three families of adrenergic receptors: $\alpha_1,\alpha_2$, and $\beta$ \cite{bylund1994international}. Stimulation of $\alpha_1$-adrenergic receptors on VTA dopaminergic neurons \cite{paladini2004noradrenergic} or those in the prefrontal cortex \cite{blanc1994blockade} promotes activity of dopaminergic neurons in the VTA. Preclinical studies found that the noradrenergic system plays role in mediating stress-induced reinstatement of cocaine seeking. Experiments on rats found that both $\alpha_2$ receptors agonists and $\beta_1$- and $\beta_2$-adrenergic  receptor antagonists can reduce stress-induced cocaine seeking behavior \cite{erb2000alpha,leri2002blockade}. Some clinical studies suggest that adrenergic blockers are effective in cocaine dependence treatment in patients with severe cocaine withdrawal symptoms \cite{kampman2006double} and are useful in reducing cocaine self-administration \cite{sofuoglu2000carvedilol}. Serotonin and norepinephrine reuptake inhibitor (SNRI) administration proved effective in attenuating cue-induced relapse to cocaine seeking after abstinence in rodents \cite{economidou2011selective}. Clinical studies on a small group of subjects indicated that SNRIs may have therapeutic potential for cocaine dependence treatment \cite{szerman2005reboxetine}. These studies indicate the promise of mediating noradrenergic signaling for the treatment for cocaine dependence.

DAT, SERT, and NET are targeted by cocaine. However, cocaine addiction involves many more proteins in their interactome with complicated molecular and functional interactions as well as a significant number of proteins upstream and downstream. The three transporters or related receptors in their protein-protein interaction networks are of frequent focus for the therapeutic treatment of stimulant dependence with the goals of an initial period of abstinence and reduced incidence of relapse \cite{kampman2005new}. On the one hand, inhibitors of the proteins could be potential therapies or assist with abstinence. The side effects and addictive liability from such inhibitors in the interactome network should be a concern. This motivated us to carry out an interactome-informed systematic investigation of the potency and off-target side effects of inhibitors pertaining to specific protein targets. A priority concern is potential effects on the human ether-a-go-go (\textit{hERG}) potassium channel, and the FDA included \textit{hERG} side effect in the most recent regulations \cite{food2005international}. 

Protein-protein interaction (PPI) networks involve both direct (physical/chemical) and indirect (functional) interactions and associations \cite{szklarczyk2019string} where a connection represents two proteins jointly contributing to a specific biological function even without direct physical/chemical interaction. Interactome provides a large number of proteins associated with a specific disease, facilitates the understanding of pathogenic mechanisms underlying the cause and progression of diseases, and promotes development of novel disease treatments. The analysis of PPIs on the proteome scale is a promising approach for the design of novel treatments for cocaine addiction as well as potential side-effects. Many interacting proteins in the interactome are collected in the String database \cite{szklarczyk2019string}, which makes it possible for interactome-informed systematic analysis of compounds targeting pathways related to cocaine addiction.

In our previous work \cite{gao2021proteome}, a machine learning/deep learning (ML/DL) approach was built around a PPI network extracted from the String dataset and centered on DAT, providing cross-target binding prediction and searching for potential inhibitors to treat cocaine addiction. ML/DL models are especially useful to predict binding affinities to targets in large scale PPI network analysis, in contrast to traditional in \textit{in vivo} or \textit{in vitro} experiments, which can be time consuming, expensive and include animal tests with ethical concerns.


In the present work, we extend our earlier effort on DAT \cite{gao2021proteome} to the SERT and NET networks to carry out systematic analysis on potential therapeutic compounds, side effects, and repurposing potentials. More extensive and comprehensive investigations could be implemented by combining the SERT, NET, and previously studied DAT networks. Special attention to the repurposing potentials of compounds targeting DAT, SERT and NET, as well as off-target side effects of inhibitors of these three transporters in the com- bined networks is considered. Potency, side effects, pharmacokinetic properties, and synthetic accessibility through ML/DL predictions form a series of filters for screening are evaluated for potential lead compounds.

\section{Methods}

\subsection{The cocaine addiction PPI networks } \label{section:ppi-network}

\begin{figure}[!ht]
	\centering
	\includegraphics[width=1.0\linewidth]{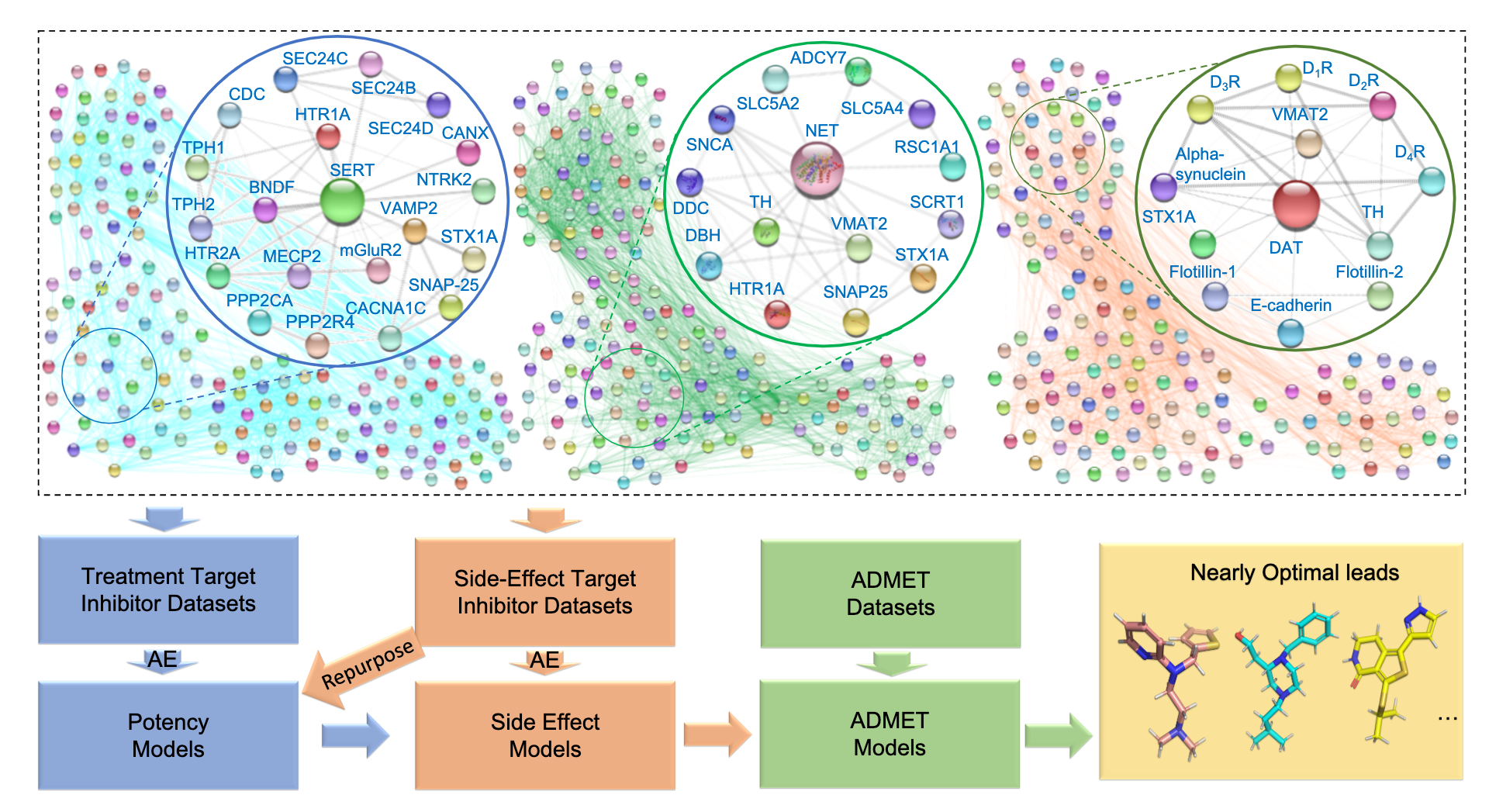} 
	\caption{ A core and global network centered on DAT, SERT and NET as well as the proteome-informed ML workflow of anti-cocaine addiction drug discovery. An auto-encoder (AE) based machine learning (ML) approach is used to encode inhibitors of proteins in the networks, and ML models are built to predict binding affinities to each protein. Screening of DAT, SERT or NET inhibitor datasets and repurposing of inhibitors from other protein targets are two key processes for drug discovery. ADMET screening is carried out following the screening or repurposing process, resulting in potentially nearly optimal leads. Abbreviations for the core SERT network: SERT (serotonin transporter ), HTR1A (5-hydroxytryptamine receptor 2A), HTR2A (5-hydroxytryptamine receptor 2A), BDNF (Brain-derived neurotrophic factor ), CANX (calnexin), PPP2R4 (Serine/threonine-protein phosphatase 2A activator), PPP2CA (serine/threonine-protein phosphatase 2A catalytic subunit alpha isoform), SEC24B (protein transport protein Sec24B), SEC24C (protein transport protein Sec24C), STX1A(syntaxin-1A ), TPH1 (tryptophan 5-hydroxylase 1), TPH2 (tryptophan 5-hydroxylase 2), DDC (dopa decarboxylase), mGluR2 (metabotropic glutamate receptor 2) and SNAP-25 (synaptosomal-Associated Protein, 25kDa). Abbreviations for the core NET network: NET (sodium-dependent noradrenaline transporter), VMAT2 (synaptic vesicular amine transporter), ADCY7 (adenylate cyclase type 7), RSC1A1 (regulatory solute carrier protein family 1 member 1), SCRT1 (transcriptional repressor scratch 1), SLC5A2 (sodium/glucose cotransporter 2), SLC5A4 (solute carrier family 5 member 4), SNCA (alpha-synuclein), TH (tyrosine 3-monooxygenase), DBH (dopamine beta-hydroxylase). Abbreviations for the core DAT network: DAT (dopamine transporter), $\rm D_{1}R $ (dopamine receptor 1), $\rm D_{2}R $ (dopamine receptor 2), $\rm D_{3}R $ (dopamine receptor 3).}
	\label{fig:networks-workflow}
\end{figure}

In addition to DAT, both SERT and NET play modulatory roles in behavioral responses to cocaine, and their functions may be pivotal to addiction. They have been frequently investigated to understand the neurobiology of cocaine addiction, providing insights for therapeutical intervention for the treatment of cocaine addiction. PPI networks help to reveal molecular interactions and cellular mechanisms. We extracted the PPI networks of SERT and NET by respectively inputting the protein names "serotonin transporter" and "norepinephrine transporter" in the String website (https://string-db.org/).

Two global networks centered on SERT and NET were extracted from String database \cite{szklarczyk2019string} for proteins that have close direct or indirect interactions with SERT or NET. The global network for SERT is formed of 151 nodes and 1,720 edges, among which a core network with 20 protein nodes that have direct interaction with SERT protein and total 66 edges are considered. The global network for NET is composed of 158 nodes, 1,791 edges and contains a core network of 14 nodes and 37 edges.

There are 20 proteins in the core  of the SERT network comprised of SERT itself and 19 other proteins that have direct interactions with SERT. SERT transports serotonin molecules from synaptic cleft back into the pre-synaptic neuron for re-packaging and re-release. It plays critical role in mediating the availability of serotonin for other receptors in the serotonergic system and terminates the effects of serotonin by removing it from the synaptic cleft. It is one of the three direct targets of cocaine, and its inhibition by cocaine may contribute to cocaine dependence. 5-HT1A receptor (5-hydroxytryptamine receptor 1A) and 5-HT2A receptor (5-hydroxytryptamine receptor 2A) are inhibitory G-protein coupled receptors for sero- tonin. Through binding to serotonin, they mediate hyperpolarisation and reduction of the firing rate of the postsynaptic neuron. Antagonists targeting 5-HT1A receptor and 5-HT2A receptor were found to diminish motivational effects of cocaine \cite{burmeister2004differential,filip2005role}. BDNF (brain-derived neurotrophic factor) promotes the survival and differentiation of selected neuronal populations of the central and peripheral nervous systems. CANX (calnexin) assists in protein assembly and quality control of the endoplasmic reticulum. PPP2R4 (serine/threonine-protein phosphatase 2A activator) accelerates the folding of proteins and acts as a regulatory subunit for serine/threonine-protein phosphatase 2A, modulating its activity or substrate specificity. PPP2CA (serine/threonine-protein phosphatase 2A catalytic subunit alpha isoform) is an enzyme providing negative control of cell growth and division. Proteins SEC24B (protein transport protein Sec24B), SEC24C (protein transport protein Sec24C) and SEC24D (protein transport protein Sec24D) are in- volved in vesicle trafficking and promoting the formation of transport vesicles from the endoplasmic retic- ulum. STX1A (syntaxin-1A) is critical for hormone and neurotransmitter exocytosis and is implicated in the docking of synaptic vesicles with the presynaptic plasma membrane. TPH1 (Tryptophan 5-hydroxylase 1) and TPH2 (tryptophan 5-hydroxylase 2) belong to the biopterin-dependent aromatic amino acid hydroxylase family and are rate limiting steps in serotonin synthesis. DDC (dopa decarboxylase) catalyzes the decarboxylation of L-3,4-dihydroxyphenylalanine (DOPA) to dopamine, L-5-hydroxytryptophan to serotonin, and L-tryptophan to tryptamine, hence modulating the amount of serotonin and catecholamine in the human body. MECP2 is widely found in neurons in the brain and promotes the maturation of the central nervous system. mGluR2 (metabotropic glutamate receptor 2) inhibits the emptying of vesicular contents at the presynaptic terminal of glutamatergic neurons. SNAP-25 (synaptosomal-associated protein, 25kDa) is critical for synaptic membrane fusion of vesicles containing neurotransmitters, as is VAMP2, and both are critical for neurotransmitter release.

The core of the NET network contains 14 proteins. NET is responsible for the reuptake of extracellular norepinephrine, extracellular dopamine and regulates the concentration of these two neurotransmitters in the synaptic cleft. VMAT2 (synaptic vesicular amine transporter) transports monoamines, especially neurotransmitters such as dopamine, norepinephrine, serotonin, and histamine, from the cytosol into synaptic vesicles. ADCY7 (adenylate cyclase type 7) is an enzyme that can catalyze the formation of cyclic AMP from ATP. RSC1A1 (regulatory solute carrier protein family 1 member 1) mediates transcriptional and post-transcriptional regulation of gene SLC5A1 and is also involved in transcrip- tional regulation of SLC22A2. SCRT1 (transcriptional repressor scratch 1) modulates the function of multiple transcription factors to regulate neuronal differentiation. SLC5A2 (sodium/glucose cotransporter 2) is the sodium-dependent glucose transporter and is responsible for the reabsorption of 80-90$\%$ of the glucose filtered by the kidney glomerulus. SLC5A4 (solute carrier family 5 member 4) may function as a glucose sensor. SNCA (alpha-synuclein) is a neuronal protein that regulates synaptic vesicle trafficking and subsequent neurotransmitter release including dopamine release and transport. TH (tyrosine 3- monooxygenase) is the enzyme responsible for catalyzing the conversion of the amino acid L-tyrosine to L-3,4-dihydroxyphenylalanine (L- DOPA), a precursor for dopamine, and its activity may be increased by cocaine exposure. DBH (dopamine beta-hydroxylase) catalyzes the conversion of dopamine to norepinephrine. SNAP-25, DDC, HTR1A and STX1A are also in the core  of the NET network and are described above.

DAT, SERT and NET constitute important components in cocaine addiction networks as they are directly inhibited by cocaine. Networks centered at the three proteins have intricate intra-network and inter-network interactions. As shown in Figure \ref{fig:networks-workflow}, the core networks formed around the three proteins serve as the center of their global networks, respectively. The SERT global network consists of four clusters of proteins,  each cluster potentially serving different functions in the serotonin system and modulatory effects of serotonin in cocaine responses. SERT and most proteins in its core network are in one of the four clusters. Proteins in the four clusters have close interactions with SERT or proteins in its core network. Both global networks for NET and DAT consist of three clusters of proteins, with most of the proteins in the core network sitting in one of the clusters. The three global networks are not independent of each other. There are many common proteins between networks. For instance, syntaxin-1A is found in all the three core networks. As discussed, SERT and NET core networks share up to four proteins. Common proteins can also be found by other pairwise network comparisons. Due to the critical roles of DAT, SERT and NET in the global protein network, and their essential modulatory effects in cocaine addiction, medications targeting them could achieve profound pharmacological effects in reducing cocaine addiction.

\subsection{Datasets}

We collected inhibitor data sets for the proteins in SERT and NET networks from ChEMBL \cite{gaulton2012chembl}. We then built models for sets with sufficient data (i.e., over 250 data points), and 37 distinct target proteins from SERT and NET networks. The details of these inhibitor datasets are listed in supporting information. Due to the role of protein \textit{hERG} in causing serious side effects in drug design, we also built a model for the \textit{hERG} inhibitor dataset.

These models allow us to carry out cross-target BA prediction for other datasets. In addition to the aforementioned 37 datasets, we also included 30 protein datasets and corresponding reliable models from the previous work \cite{gao2021proteome} for the DAT network, as they may be involved in cocaine addiction. The DAT network shares some common proteins with the current SERT-NET network. In total, we built 61 different datasets with models.

\subsection{The seq2seq LV-FP}

Molecular fingerprints are used to represent molecules, usually in the form of vectors with each vector element indicating the existence, degree, or frequency of each structural characteristic or property. These fingerprints have a variety of applications in ML/DL analysis, virtual screening, similarity-based compound searches, target molecule ranking, drug ADMET prediction, and other drug discovery processes \cite{gao20202d}.

2D fingerprints, mathematical representations, and latent-vector fingerprints (LV-FPs) constitute three types of molecular fingerprints frequently used in ML/DL models. Molecular feature extraction by means of state-of-the-art 2D fingerprints can be classified into four categories including substructure key-based fingerprints, topological or path-based fingerprints, circular fingerprints, and pharmacophore fingerprints \cite{gao20202d}. MACCS, FP2, ECFP and 2D-pharmacophores are all popular 2D fingerprints and can be generated by the open-source cheminformatics software RDKit. In addition, models based on deep neural network (DNN)  draw interest in extracting mathematical features. Many differential geometry, algebraic topology, and spectral graph-based approaches have been investigated in this regard \cite{nguyen2020review}. LV-FPs are  the molecular representations in the neutral layers of DNN architectures and gained wide popularity in a drug discovery \cite{winter2019efficient,gao2020generative} and molecular analysis \cite{xu2017seq2seq,wu2018moleculenet}. Among those DNN architecture models, the seq2seq model draws particular attention. In this work, we adopted the LV-FP of molecules generated by our in-house seq2seq models.

The seq2seq model is a DL autoencoder architecture originated from natural language processing. It has achieved breakthrough success in English-French translation and conversational modeling. The seq2seq model starts with mapping an input sequence to a fixed-sized latent vector in the latent space using a gated recurrent unit (GRU) \cite{cho2014learning} or a long short-term memory (LSTM) network \cite{hochreiter1997long}, and then maps the vector to a target sequence with another GRU or LSTM network. These vectors are called latent vectors and are often used to characterize the source or target sequence. 

In our study, input and output sequences are both SMILES strings -- a one-dimensional ``language'' of chemical structures. Using nearly 104 million molecules, our autoencoder model is trained to have a high reconstruction ratio between input and output smiles strings while latent vectors carry faithful information of the chemical structures. These LV-FPs can be used to represent compounds. The seq2seq model and LV-FPs were realized by our in-house source code. We applied bidirectional LSTMs as the encoder. The generated LV-FPs have a dimension of 512.

\subsection{BA predictions}

The features encoded with LV-FPs allow us to build accurate machine learning models. In our implementation, gradient boosting decision tree (GBDT) regressor in scikit-learn (version 0.20.1) \cite{pedregosa2011scikit} was adopted to build models, and we then use these models to make BA predictions. In our study of cocaine addiction, we built models for the 61 datasets. The hyperparameters were tuned according to 10-fold cross-validation tested on each dataset, and they are summarized in supporting information. Finally, we validated the effectiveness and robustness of our models by comparisons with the literature in our previous work \cite{gao2021proteome}.

\section{Results}

\begin{figure}[!ht]
	\centering
	\includegraphics[width=0.97\linewidth]{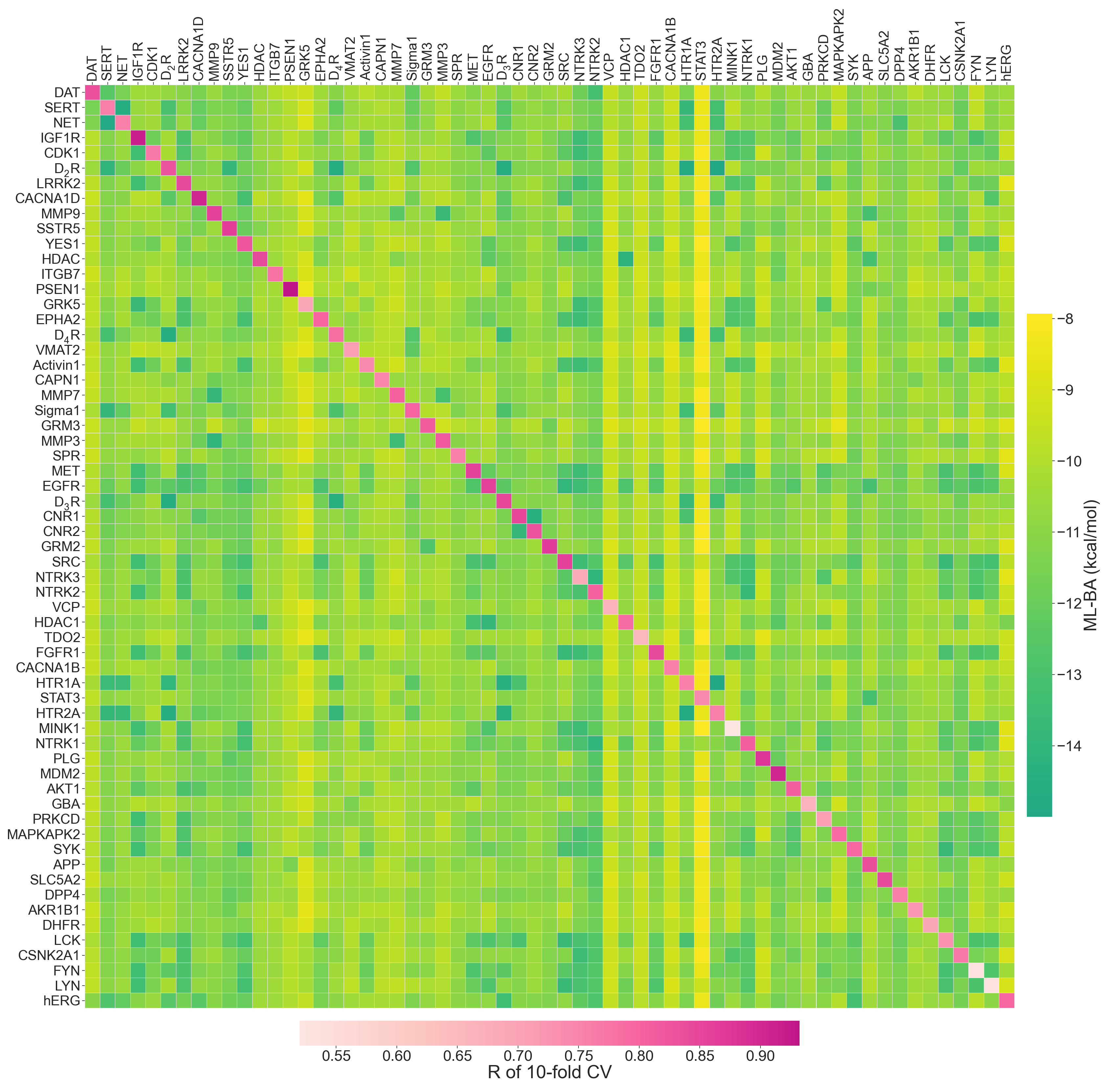} 
	\caption{Heatmap of cross-target BA prediction indicating the inhibitor specificity of each dataset. In each row, the diagonal element shows the Pearson correlation coefficients of 10-fold cross-validation (R of 10-fold CV) on the machine-learning predicted BAs (ML-BAs) of each dataset. Off-diagonal elements represent the highest ML-BAs of the inhibitors in each dataset to other targets.}
	\label{fig:cross-prediction-heatmap}
\end{figure}

\subsection{Reliability tests}

 In this section, we adopt ML/DL models to systematically predict inhibitor BAs to analyze side effects and repurposing potential. 

These ML models show promising predictions with Pearson's coefficient (R) in 10-fold cross-validation (CV) tests for each inhibitor dataset as indicated in the heatmap of Figure \ref{fig:cross-prediction-heatmap}. 11 of the 38 new models for inhibitor datasets have R greater than 0.8, and 28 of the 38 models have R$>$0.7. Only 3 of the 38 have R lower than 0.6. The details of the new 38 models in addition to previously built DAT models can be found in supporting information.

\subsection{Cross-target BA predictions} 
The highest BA of cross-target prediction represents the side-effect strength of inhibitors to other targets. The 3,660 cross-target predictions in Figure \ref{fig:cross-prediction-heatmap} exhibit 3,338 potential side effects determined by BAs higher than -9.54 kcal/mol ($\rm K_i=0.1 \mu$M). On the other hand, the remaining 322 predictions with all BA values greater than -9.54 kcal/mol suggest weak side effects. Figure \ref{fig:cross-prediction-heatmap} depicts the results of the cross-target BA prediction of the 61 inhibitor datasets. Along the diagonal line are the R values of 10-fold CV tests for the corresponding protein datasets, while the off-diagonal elements indicate the predicted minimal BA of the datasets to a specific target protein. For example, the $j$-th element in the $i$-th row shows the predicted highest BA of $i$-th inhibitor dataset listed to left of the heatmap by the $j$-th target model listed on top of the heatmap.

\subsubsection{Cross-target BA correlation revealing binding site similarities}

\begin{figure}[!ht]
	\centering
	\includegraphics[width=0.95\linewidth]{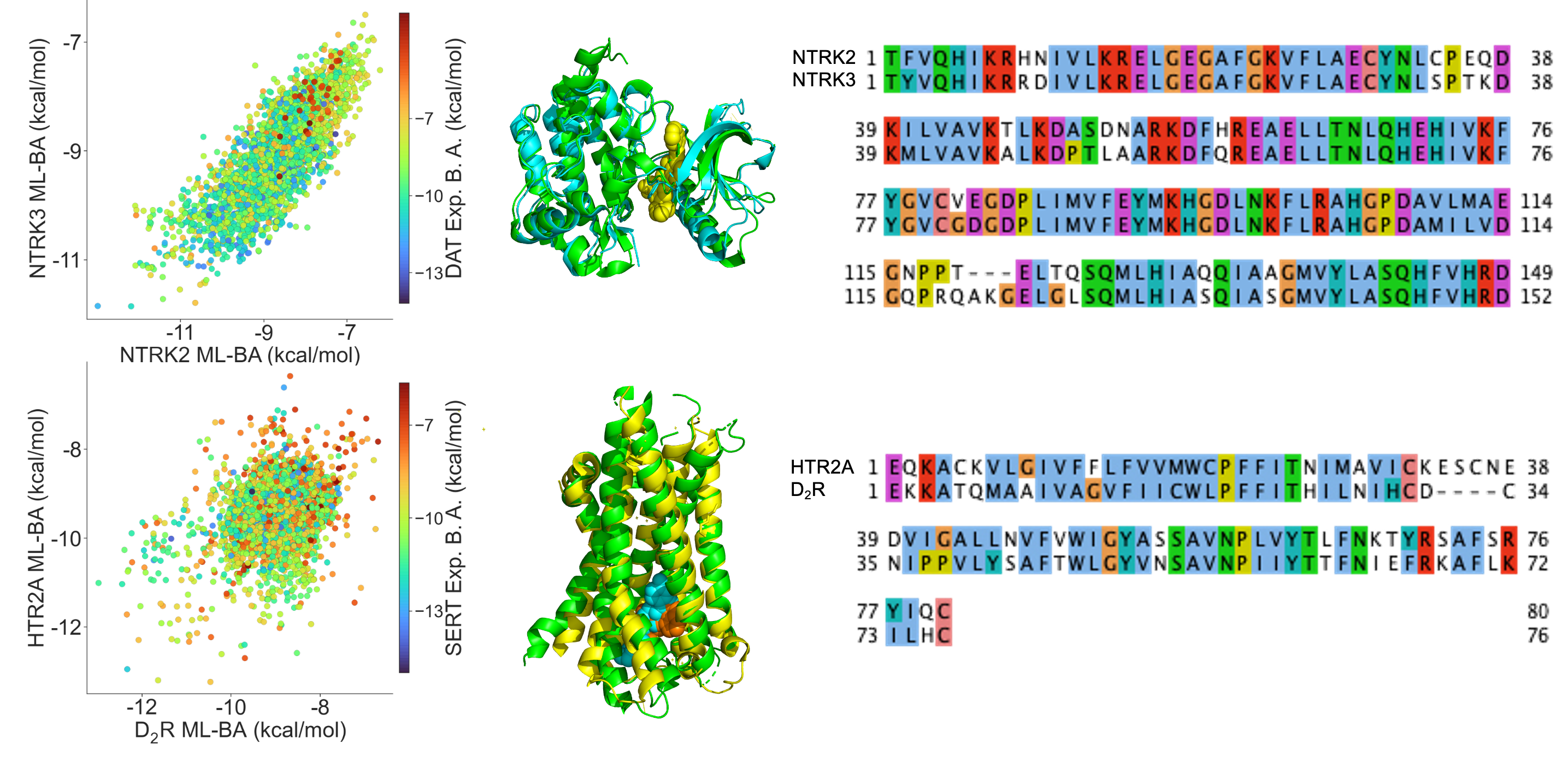} 
	\caption{Examples of cross-target predicted BA correlations detecting binding site similarities of different proteins. The first column exhibits the predicted BAs of the inhibitor dataset to different targets, the second column corresponds to the 3D structure alignment of two proteins, and the third column displays the 2D sequence alignment of the binding site. The PDB IDs are 4AT4, 6KZC, 6CM4 and 6WGT for NTRK2, NTRK3, $\rm D_{2}R$ and HTR2A, respectively.}
	\label{fig:2Dalign-paper}
\end{figure}

Binding site similarities can yield cross-target BA correlation. On the other hand, high BA correlation can help identify binding site similarities. According to our cross-target prediction, there are some examples of high BA correlation to similar binding sites. For instance, NTRK2 (Tropomyosin receptor kinase B) and NTRK3 (Tropomyosin receptor kinase C) are all in the family of tropomyosin receptor kinase Trk and are activated by protein nerve growth factors or neurotrophins. As shown in the first row of Figure \ref{fig:2Dalign-paper}, our ML-models predict that the BAs of DAT inhibitor dataset to NTRK2 and NTRK3 have a Pearson correlation coefficient (R) of 0.79. Their 3D protein structure alignment displayed on the right shows their highly similar structure. And the 2D sequence alignments of the binding sites with sequence identity $84.53\%$ confirm their biding site similarity.

We also found that the proteins from different families can also have similar binding sites, leading to cross-target BA correlations, and an example is shown in the second row of Figure \ref{fig:2Dalign-paper}. The predicted BAs of the SERT inhibitor dataset to $\rm D_2R$ and HTR2A have a Pearson correlation coefficient (R) of 0.31. $\rm D_2R$ and HTR2A are receptors for dopamine and serotonin, respectively, but they are not in the same family. The highly similar 3D structure is shown on the right, and the 2D sequence alignments of the binding site have a sequence identity of $42.11\%$. Some additional examples of similar binding sites can be found in supporting information. In addition, the pairwise full sequence similarities of 60 proteins are compared in our study, and the identity scores are shown in Figure S3. High full sequence similarities could also indicate local binding site similarities. The full sequence identity of $\rm NTRK2$ and $\rm NTRK3$ is as high as $63.2\%$, which is not far from the local binding site similarity score $84.53\%$. Our predicted BA correlation could be a better tool in detecting similarities, since the high similarity of the local binding site could directly lead to high BA corrections.

\subsubsection{Predictions of side effects and repurposing potentials}

It is desirable for a drug candidate to be highly specific, i.e., having a high BA to its target and very low BAs to all other human proteins, reducing likelihood of side effects. Additionally, if a drug candidate interacts weakly with its designated target but is potent to another unintended protein, it has repurposing potential. Our ML model is a useful tool to study possible side effects and repurposing potentials by systematically carrying out cross-target predictions. We adopted the 61 available models to predict the BAs of compounds in other datasets to specific protein targets. The 61 inhibitor datasets and related models allowed us to evaluate possible side effects of drug candidates as well as their repurposing potentials for other protein targets.

\begin{figure}[ht!]
	\centering
	\includegraphics[width=1.0\textwidth]{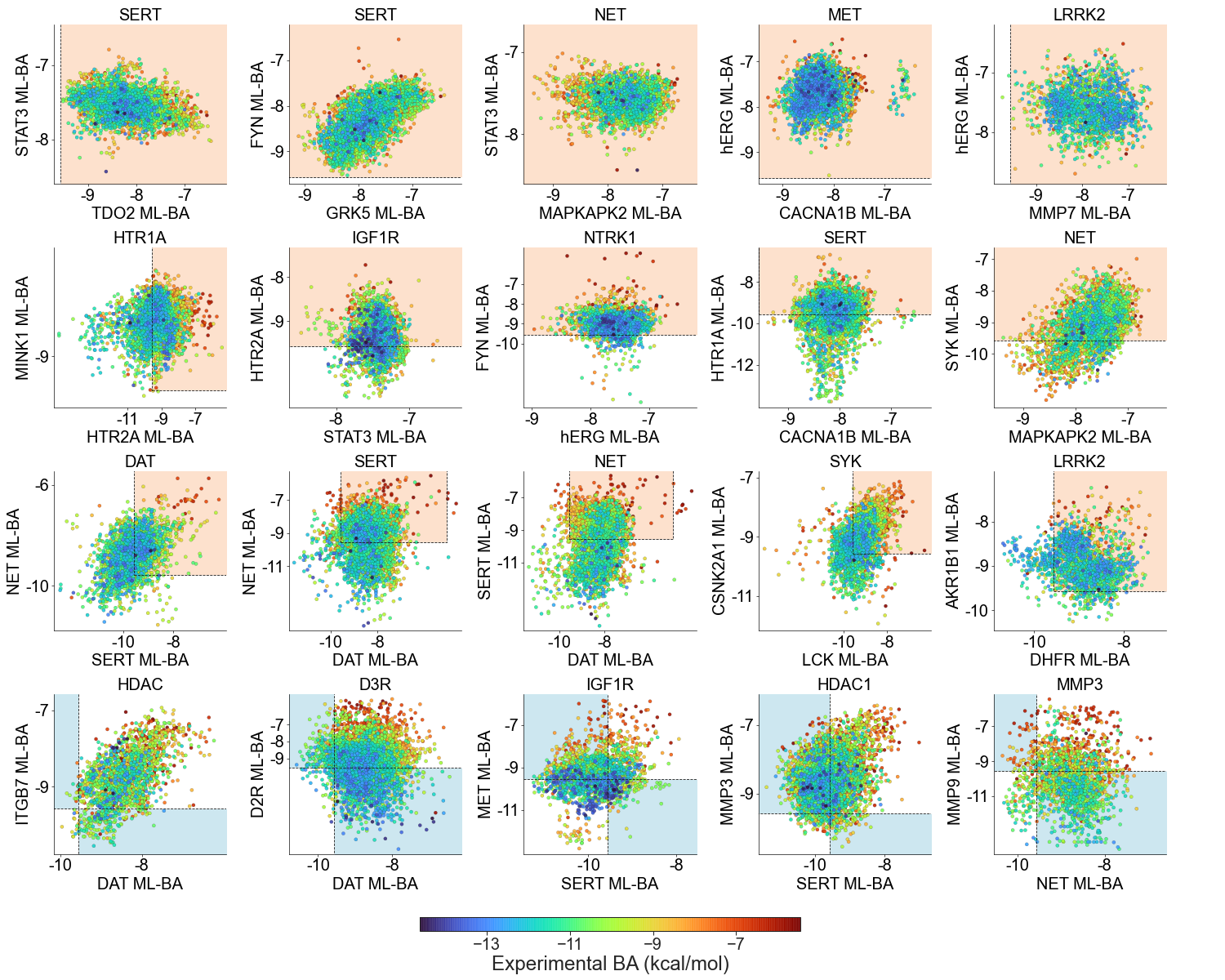}
	\caption{Examples of inhibitors' possible side effects or repurposing potentials. The first three rows list some inhibitor datasets that have side effects to 0, 1 or 2 of two off-target proteins. The orange rectangular frames outline the ranges where no side effects of inhibitors are caused to either off-target protein. The last row displays some inhibitor datasets that have repurposing potential to other proteins. The two blue rectangles highlight the domains where inhibitors can have repurposing potential to one protein but have no side effect to the other one}
	\label{fig:SideEffect-Repurposing-selected}
\end{figure}

Figure \ref{fig:SideEffect-Repurposing-selected} plots some examples of inhibitors' side effects and repurposing potentials by prediction. Each chart involves one inhibitor dataset and two target proteins. The $x$ and $y$ values represent the predicted BAs of a designated inhibitor dataset to two other proteins, and the color of dots indicates the experimental BA of inhibitors to designated proteins: blue color represents strong BAs, and red color denotes low BAs. The first three rows depict cases of predicted side effects, and the last row depicts cases of repurposing potential. The orange frames in the first three rows highlight the domains where inhibitors for a designated protein would not have side effects on the other two targets,  i.e., BAs $>$ -9.54 kcal/mol ($\rm K_i$=0.1 $\rm \mu$M). The two blue frames in each chart of the last row indicate the ranges where inhibitors have repurposing potential to one target, i.e. predicted $<$ -9.54 kcal/mol ($\rm K_i$=0.1 $\rm \mu$M)) and no side effects are caused on the other one, i.e. predicted BA $>$ -9.54 kcal/mol.

The first row of Figure \ref{fig:SideEffect-Repurposing-selected} shows cases of inhibitor dataset causing no side effects to two other targets. All the active inhibitors of desired targets are predicted to have low BAs to other targets. For example, the first and second charts indicate that all active inhibitors of SERT are predicted to cause no side effect to TDO2, STAT3, GRK5, and FYN proteins. It is the same case in the third and fourth charts that NET inhibitors have no predicted side effects on MAPKAPK2, STAT3, CACNA1B, and \textit{hERG} proteins. The second and third rows show inhibitors of the designated protein having side effects on either or both of two other targets. The first panel in the second row shows that the HTR1A dataset has more than half of its inhibitors predicted to bind with affinity $<$ -9.54 kcal/mol to HTR2A. This can be due to high structural similarity as they are both serotonin receptors. The first three charts in the third-row show that a large number of inhibitors of DAT, SERT, and NET can cause side effects to two other proteins.

Some examples of inhibitors with repurposing potentials are provided in the last row of Figure \ref{fig:SideEffect-Repurposing-selected}. The third panel in the fourth row shows that many inactive inhibitors of IGFR1 may have repurposing potentials to SERT and have no side effect on MET. In the fifth panel of the fourth row, we can observe that several inactive inhibitors of MMP3 may have repurposing potentials for NET and MMP9.

\subsubsection{Repurposing potential for SERT, NET, and DAT and side effects on \textit{hERG}}

DAT is a well-known mediator of cocaine’s behavioral effects. SERT and NET also play significant roles in cocaine responses and inhibitors of each have been considered as potential medications for cocaine dependence. In this vein, we carried out BA cross predictions of inhibitors from other datasets in the PPI networks and the \textit{hERG} inhibitor dataset to find compounds with repurposing potentials for either SERT or NET. As discussed above, compounds with repurposing potential should be inactive to their designated target with experimental BA value $>$ -9.54 kcal/mol yet having BA values $<$ -9.54 kcal/mol for SERT or NET. Since \textit{hERG} is a priority side effect concern for novel medications, the side effect threshold to \textit{hERG} was considered -8.18 kcal/mol ($\rm K_i$=1 $\rm \mu$M).

Figures S11 and S12 in the supporting information show the predicted repurposing potentials to SERT and potential \textit{hERG} effects of inhibitors from 59 other datasets. We show that a larger portion of the datasets may be qualified for further screening. In some datasets, nearly half of all compounds may have repurposing potential and low \textit{hERG} effects. In the CDK1 dataset, 534 out of 1253 inhibitor compounds possessed repurposing potential, and 507 still remain after considering \textit{hERG} side effects. Among the 392 DHFR inhibitors with repurposing potentials to SERT, 349 had no predicted \textit{hERG} side effect. In addition, DAT and NET datasets provided 685 and 451 such compounds, respectively.

Predictions of repurposing potentials and \textit{hERG} side effects of 59 datasets for NET are provided in Figures S13 and S14. A fairly large number of compounds were obtained from the 59 datasets according to the prediction. For example, 151, 155, 229, and 382 compounds with repurposing potential to NET and low predicted \textit{hERG} side effects were found in the DHFR, LCK, DAT and SERT datasets, respectively.

The predicted repurposing potential for DAT and \textit{hERG} side effect are shown in Figure S15 and S16 in the supporting information. Here, we only consider inhibitor datasets in the SERT-NET networks yet not in the DAT network. The predictions for the datasets in the DAT network were already reported in our previous work \cite{gao2021proteome}. Most of the datasets have a large portion of the inhibitors whose predicted \textit{hERG} BA values are greater than -8.18 kcal/mol, suggesting no serious potential \textit{hERG} side effect. According to our predictions, several datasets including DPP4, FGFR1, HDAC1, HTR1A, HTR2A, HDM2, NET, SERT and SRC have around half of their inhibitors with high \textit{hERG} side effects. On the other hand, in search for compounds with repurposing potential to DAT, most datasets have a limited number of compounds satisfying repurposing requirements. However, a few datasets contain several compounds with repurposing potential and low predicted \textit{hERG} side effects. For instance, the LCK dataset of 1855 inhibitors contains 17 compounds with repurposing potential to DAT. Fortunately, all 17 compounds are predicted to have no \textit{hERG} side effects, with predicted BA values to \textit{hERG} $>$-8.18 kcal/mol. The SERT dataset has 88 compounds that are predicted to have repurposing potential to DAT. 41 of the 88 compounds are predicted to have no \textit{hERG} risk. In the NET inhibitor dataset, 43 out of the 75 compounds with repurposing potential have low predicted \textit{hERG} potential. In summary, an encouraging number of inhibitor compounds with repurposing potential to DAT and low predicted \textit{hERG} side effects are available from the various datasets in the three networks. According to predictions, 41, 43, 11, 35 and 19 compounds can be obtained respectively from SERT, NET, DPP4, HTR1A and HTR2A inhibitor datasets.

\subsubsection{Possible Side effects of SERT, NET, and DAT inhibitors to other proteins}

Herein, we investigate the possible side effects of inhibitors of SERT, NET, and DAT. Figures S8, S9, and S10 in the supporting information show the predicted side effects of SERT, NET and DAT inhibitor datasets, respectively.

Figures S8 and S9 show the BA predictions of inhibitors of SERT and NET to other proteins. It can be seen that nearly half of the SERT inhibitors may have side effects on $\rm D_3R$ and YES1 proteins in the DAT network. This convinced us to carry to complete investigations across networks. Many NET inhibitors risk side effects from several targets including LRRK2, Sigma1, SERT, HTR2A, $\rm D_4R$ and others.

Figure S10 shows that DAT inhibitors could cause side effects on several proteins including GRK5 and STAT3 from BA predictions. In addition, many DAT inhibitors may cause serious side effects through other proteins including HTR1A, HTR2A, $\rm D_2R$, $\rm D_3R$, SERT, NET, and others. HTR1A and HTR2A serve as serotonin receptors, while $\rm D_2R$, $\rm D_3R$ play roles as dopamine receptors. Such side effects of DAT inhibitors on the receptors may interfere with dopamine or serotonin transmission. HRT1A and HTR2A do not belong to the network of DAT, demonstrating the likelihood that a DAT inhibitor with no side effects in its own network can pose serious side effects on proteins in other networks. Thus, it is reasonable to consider the side effects of DAT inhibitors on a larger scale involving more proteins.

\subsection{Druggable property screening}

\begin{figure}[ht!]
	\centering
	\includegraphics[width=1.0\textwidth]{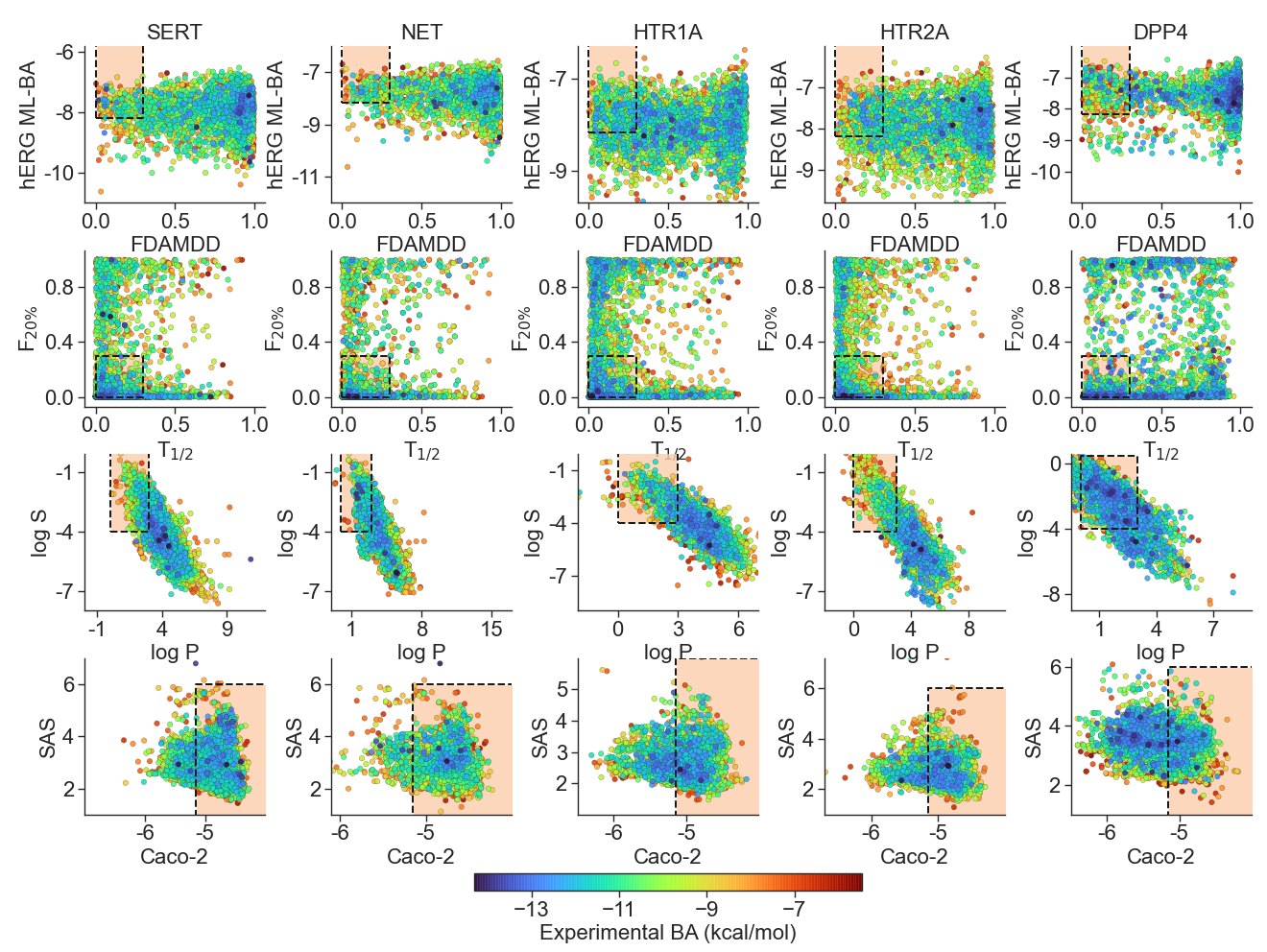}
	\caption{Druggable property screening based on ADMET properties, synthesizability, and \textit{hERG} side-effects to compounds from five critical protein datasets: SERT, NET, HTR1A, HTR2A, DPP4. The colors of points represent the experimental BAs to these targets. The $x-$ and $y-$axes show predicted ADMET properties, synthesizability, or \textit{hERG} side-effects. Orange frames outline the optimal ranges of these properties and side effects.}
	\label{fig:admet-dist}
\end{figure}

\begin{table}
	\centering
	\begin{tabular}{c|c}		
		\hline
		Property & Optimal range   \\ 	
		\hline
		FDAMDD & Excellent: 0-0.3; medium: 0.3-0.7; poor: 0.7-1.0  \\
		$\rm F_{20\%}$ & Excellent: 0-0.3; medium: 0.3-0.7; poor: 0.7-1.0  \\
		Log P & The proper range: 0-3 log mol/L \\
		Log S & The proper range: -4-0.5 log mol/L \\
		$\rm T_{1/2}$ & Excellent: 0-0.3; medium: 0.3-0.7; poor: 0.7-1.0  \\
		Caco-2 & The proper range: $>$-5.15 \\
		SAS & The proper range: $<$6 \\
		\hline
	\end{tabular}
	\caption{The optimal ranges of  6 selected ADMET characteristics and synthesizability (SAS) were considered in this work.}
	\label{tab:property-optimal}
\end{table}

We performed systematic screenings on ADMET properties, synthetic assessibility (SAS), and \textit{hERG} risk of all inhibitor datasets. Accurate predictions of pharmacokinetic properties are vital for drug design. ADMET (absorption, distribution, metabolism, excretion, and toxicity) includes a diversity of attributes associated with the pharmacokinetic attributes of a compound, and is an important factor in drug design \cite{wager2016central}. In this work, we restrict our attention to six indexes of ADMET, i.e., FDAMDD, $\rm T_{1/2}$ and $\rm F_{20\%}$, log P, log S, and Caco-2, and SAS \cite{gao2021proteome}, as well as a \textit{hERG} risk assessment. The optimal ranges of ADMET properties and SAS are provided in table \ref{tab:property-optimal}, while the BA value $>$ -8.18 kcal/mol is applied as the required range for exempting \textit{hERG} side effects. The available ML models for ADMET \cite{xiong2021admetlab} and SAS as well as our models for cross-target prediction enable us to systematically search for promising compound leads with desired ADMET properties.

Figure \ref{fig:admet-dist} illustrates an example of screening datasets of five proteins: SERT, NET, HTR1A, HTR2A, and DPP4. SERT and NET are two of the direct targets of cocaine, and efforts are often dedicated to pharma- cological effects on these two proteins in cocaine treatment. Serotonin receptors including HTR1A and HTR2A also draw attention since some pharmacological manipulations were found to reduce cocaine use in preclinical studies \cite{filip2005serotonergic}. Each column records the ADMET predictions of the given inhibitor dataset, while each row in Figure \ref{fig:admet-dist} represents a pair of ADMET characteristics. The orange frames indicate the optimal domains of the specified two characteristics for the inhibitor dataset. Finally, the dot color portrays the experimental binding affinities.

The first row corresponds to the FDA maximum recommended daily dose (FDAMDDs) and the BA for \textit{hERG} (\textit{hERG} BA), which reflect toxicity of potential drug candidates to the human body. A small fraction of SERT and NET datasets lie in the optimal domains outlined by this pair of properties. The restriction by FDAMDD filters out more than two-thirds of inhibitors from all the datasets. The second row stands for the screening of absorption properties $\rm T_{1/2}$ (half-life) and $\rm F_{20\%}$ (human oral bioavailability $20\%$). The half-life is the amount of time it takes for a drug's active substance to reduce by half in human body, and $\rm T_{1/2}$ indicates the probability of half-life less than 3 hours. $\rm F_{20\%}$ represents the probability of an oral drug reaching systemic circulation with less than 20 percent of initial dose remaining. This pair of properties together put on strict thresholds for drug screening, as we can see that the orange frames only cover a small portion of a given dataset. Moreover, the third row of Figure \ref{fig:admet-dist} denotes the screening on log P and log S, which are the logarithm of aqueous solubility value and the n-octanol/water distribution coefficient, respectively. These two screening properties fail many inhibitors, especially from the NET dataset. The last row depicts caco-2 and SAS screening. Caco-2 is commonly used to estimate \textit{in vivo} permeability of oral drugs, while SAS is designed to estimate ease of synthesis of drug-like molecules. Based on the predictions, most inhibitors of the five datasets are not hard to synthesize. As a result, these two properties allow a large portion of inhibitors to pass the screening.

The ADMET and other properties are important indexes of eligible candidate drug compounds. We anticipate that several of these properties may pose significant challenges when searching for desired drugs to treat cocaine addiction. Reliable ML-based models are in need to accomplish the prediction of these properties. We took advantage of the ADMETlab 2.0 solver \cite{xiong2021admetlab} to obtain these screening results.

\section{Discussion}

\subsection{Side-effect predictions of existing experimental medications}

Due to the importance of noradrenergic and serotonergic systems in mediating cocaine effects, some experimental medications targeting these two systems have been investigated. In this study, we utilized ML models to predict the side effects of these medications.

\subsubsection{Medications targeting the serotonin system}

\begin{figure}[!ht]
	\centering
	\includegraphics[width=0.9\linewidth]{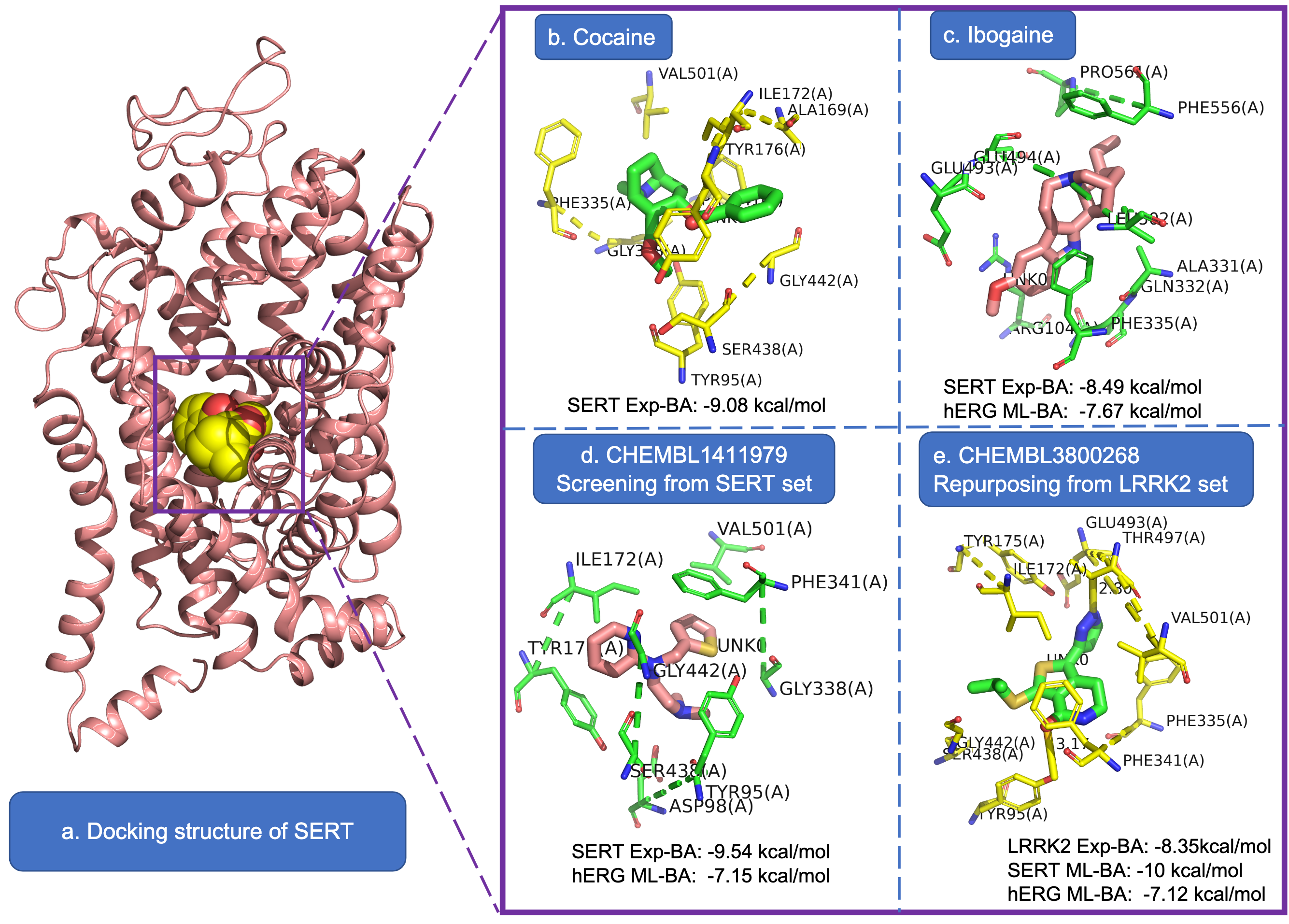} 
	\caption{The docking structure of SERT with cocaine, ibogaine, CHEMBL14111979, and CHEMBL3800268. The experimental or predicted binding affinities of cocaine, Ibogaine and two compounds from screening and repurposing are also presented.}
	\label{fig:docking}
\end{figure}

Cocaine dependence is closely linked to deficits in the serotonin system. Some preclinical studies have already shown that self-administration of cocaine can be reduced through serotonin-enhancing medications \cite{carroll1990fluoxetine,peltier1993effects}. On the other hand, pharmacologic manipulation of serotonin associated with the dopamine system \cite{filip2005serotonergic} and can indirectly modulate the dopamine circuits relevant to dependence. Serotonin-enhancing medications also benefit the dopamine system by indirectly increasing extracellular dopamine levels. Among those serotonergic medications, ibogaine is investigated most frequently. Its docking structure to SERT and some experimental/predicted BAs are shown in Figure \ref{fig:docking}c. 2D structures and experimental or predicated BA to various targets of many other experimental medications are shown our supporting information.

\paragraph{Ibogaine and its derivatives}

Ibogaine is a hallucinogenic alkaloid found in the root and bark of the African shrub, Tabernanthe iboga. It may have effects in the treatment of not only cocaine dependence but also for alcohol, opiate, and methamphetamine dependence. The pharmacology of Ibogaine is complex, and it has affinities for  $k$-opioid receptors, $N$-methhyl-D-aspartate receptors, $\sigma1$ and $\sigma2$ receptors, as well as DAT and SERT \cite{glick1998mechanisms}. The psychoactive effects of Ibogaine are associated with $k$-opioid receptors \cite{zubaran1999noribogaine}, while the agonistic action with serotonin 5-$\rm HT_{2A}$ receptor (HTR2A) may contribute to the hallucinogenic effects \cite{wei1998acute}. Its actions are complex mediating different neurotransmitters at the same time. In particular, its interaction with DAT and SERT may underlie its anti-addiction properties for cocaine and it has been promoted as a treatment for addiction in Europe and North America.

Ibogaine inhibits both DAT and SERT with$\rm IC_{50}$ values of 4.0 $\rm \mu M$ and 0.59 $\rm \mu M$, respectively \cite{efange1998modified}. The mechanism of Ibogaine inhibition of SERT is different from other known inhibitors in the sense that it is not competitive with substrate and it stabilizes the transporter in an inward-open conformation. Ibogaine binds to a site accessible from the cell exterior that does not overlap with the substrate-binding site on SERT and DAT \cite{bulling2012mechanistic}. Despite its promising effects for the treatment of cocaine addiction, its associated side effects, including death, are a serious concern, and have led to its prohibition in some countries. From 1990 to 2008, 19 fatalities associated with the ingestion of Ibogaine were reported, and 6 of these fatalities were caused by acute heart failure or cardiopulmonary arrest \cite{koenig2015anti}.

One side-effect of Ibogaine is that it may cause long QT syndrome at higher doses, perhaps by blocking \textit{hERG} potassium channel in the heart \cite{koenig2015anti}. The predicted BA value of Ibogaine to \textit{hERG} using our model is -8.43 kcal/mol, and this modest prediction can indicate the potential of cardiac risk. Additionally, our models anticipated other high risk side effects. The predicted BA value to YES1 is -9.71 kcal/mol, and YES1 inhibition is associated with  sarcoma and acute myeloid leukemia. Ibogaine is also predicted to have high BAs of -10.69 kcal/mol, -10.47 kcal/mol, -10.18 kcal/mol and -10.70 kcal/mol to NTRK1, NTRK2, NTRK3, and SYK, respectively.

In recent years, there is increased interest in 18-methoxycoronaridine (18MC), which is a derivative of ibogaine. It has shown its effectiveness in reducing self-administration of cocaine, morphine, methamphetamine, nicotine, and sucrose in preclinical models \cite{glick2000development}. It has similar pharmacological effects to those of ibogaine, but it does not cause tremors, Purkinje cell dysfunction, or toxicity in the brain \cite{glick199618}. Moreover, 18MC has no affinity for SERT, in contrast to Ibogaine, and so provides an enhanced safety profile to hu- mans compared to ibogaine and is under clinical trials. The predicted \textit{hERG} BA of 18MC by our model is -7.67 kcal/mol, which reflects moderate potential in incurring heart issues. However, it is predicted to have BA values of -10.42 kcal/mol, 10.15 kcal/mol, -10.20 kcal/mol, 10.15 kcal/mol and -9.91kcal/mol, respectively for SSTR5, YES1, LRRK2, VMAT2, and CNR2. Diseases associated with SSTR5 include acromegaly, pituitary adenoma and prolactin-pecreting. YES1 is associated with sarcoma. These strong off-target binding affinities may indicate the potential risk of such diseases or related side effects when 18MC is tested in humans.


\paragraph{Selective serotonin reuptake inhibitors}

Selective serotonin reuptake inhibitors (SSRIs) are typi- cally used as antidepressants. They increase the extracellular level of serotonin by limiting its reuptake into the presynaptic cell. Generally, SSRIs have a stronger affinity to SERT than to DAT or NET. As a consequence of their enhancement of extracellular serotonin levels, attempts have been made to use SSRIs to treat cocaine addiction, and some clinical trials using SSRIs have shown some promise in treating cocaine addiction.

Fluoxetine is an SSRI approved by FDA to treat several psychiatric disorders including depressive depression, bulimia nervosa, and others. Some preclinical studies have shown the effectiveness of fluoxetine in the treatment of cocaine addictions, but clinical trials have yielded mixed results. Some studies showed the efficacy of fluoxetine in significantly reducing cocaine use \cite{walsh1994fluoxetine,schmitz1998medication}, while others showed that fluoxetine is not effective in altering cocaine effects \cite{batki1996controlled}. However, it is encouraging that fluoxetine tends to be more effective if higher doses are used in the cocaine treatment \cite{walsh1994fluoxetine,batki1996controlled}. Our ML models show that fluoxetine has low binding affinities to most of the 61 targets in our network except for SERT, with predicted BA lower than -10 kcal/mol. The high binding affinity to SERT is reasonable since fluoxetine is an SSRI. Fluoxetine is predicted to have binding affinity higher than -9.0 kcal/mol to 55 out of the 61 targets, which reflects its low side effects on the targets in the network. Since Fluoxetine is already an FDA approved medication, such low side effects are anticipated. It is predicted to have binding affinity of -8.25 kcal/mol to \textit{hERG}, which indicates relatively low potential to cause prolongation of the QT interval.

Sertraline is also an SSRI used as antidepressant for various psychiatric conditions. It is given as a generic medication, and it was the most prescribed psychiatric medication in the USA in 2016. It has also been tested in clinical trials for the treatment of cocaine dependence. Sertraline is effective in reducing cocaine cravings and can produce delays in relapse in recently abstinent cocaine abusers with depressive symptoms. According to our ML model predictions, sertraline has low binding affinities to most of the 61 targets in the networks. It is predicted to have relatively high BAs of -9.15 kcal/mol, -9.51kcal/mol and -9.42 kcal/mol for $\rm D_4R$, SYK and LYN, respectively. LYN is associated with sarcoma and cutaneous mastocytosis. The predictions of BAs of sertraline for LYN may indicate a risk of issues related to these diseases. The predicted BA to \textit{hERG} is -7.75 kcal/mol, which suggests a low potential for causing heart issues.

Citalopram is another SSRI antidepressant. Recently, some very encouraging results were obtained in reducing cocaine use when given in combination with contingency management \cite{moeller2007citalopram}. Side effects were observed to be mild. These studies provided support that citalopram combined with behavioral therapy can be a promising treatment for cocaine dependence. According to our ML model predictions, citalopram has low BAs to most of the 61 protein targets. It is predicted to have -9.30 kcal/mol, -9.34 kcal/mol, -9.72 kcal/mol, and -10.11kcal/mol to SSTR5, $\rm D_3R$, CNR1, and HTR2A, respectively. SSTR5 is associated with diseases including acromegaly and prolactin-secreting pituitary adenoma. The binding affinity of citalopram indicates potential risk for the aforementioned diseases. Its strong binding affinity to 5-$\rm HT_{2A}$  and $\rm D_3R$ indicate its effect in the transmission of neurotransmitters serotonin and dopamine. 5-$\rm HT_{2A}$ and $\rm D_3R$ have been investigated as pharmacological targets in the treatment of cocaine dependence, and citalopram may cause unexpected effects due to binding these two targets. The predicted BA to \textit{hERG} is -8.16 kcal/mol, consistent with the announcement that " causes dose-dependent QT interval prolongation" by FDA \cite{filip2005serotonergic}.

\paragraph{5-HT3 receptor antagonists}
Efforts to target the 5-$\rm HT_{1B}$, 5-$\rm HT_{2A}$ and 5-$\rm HT_{3}$ receptors have been made due to their  important roles in potential cocaine addiction mechanisms. Ondansetron, a 5-$\rm HT_{3}$ receptor antagonist, has been investigated in clinical trials for cocaine addiction treatment. Preclinical studies showed its efficacy in abolishing the reinstatement of cocaine administration \cite{davidson2002ondansetron}, and further studies showed that ondansetron can be particularly effective in reducing oral cocaine self-administration when given during the acute cocaine withdrawal period \cite{davidson2004ondansetron}. Our ML models show its binding affinity values of -9.82 kcal/mol, -9.64kcal/mol, -9.86 kcal/mol and -9.4 kcal/mol for $\rm D_4R$, Sigma1, HTR1A, and DPP4, respectively. The high binding affinities to these targets indicate the potential for side effects mediated by them.

\subsubsection{Medications targeting the noradrenergic system}

Some preclinical and clinical studies showed that pharmacological manipulations of the noradrenergic systems could be a potential treatment for cocaine addiction \cite{sofuoglu2009norepinephrine}. In noradrenergic systems, norepinephrine (NE) is the main chemical messenger and plays a contributing role in mediating the rewarding effects of co- caine. NET is regarded as a potential target for treatment of cocaine addiction. Atomoxetine and reboxetine are two selective noradrenaline reuptake inhibitors with NET-blocking effects.

Reboxetine, an antidepressant medication, has been tested in the treatment of cocaine addiction \cite{szerman2005reboxetine} and reported as an effective and safe therapeutic option. However, more rigorous double-blind studies of reboxetine need to be carried out before its efficacy in the treatment of cocaine dependence can be fully confirmed. Another promising outcome was reported combining reboxetine with the SSRI escitalopram \cite{camarasa2005escitalopram}. Predictions from our ML models shows side effects to $\rm D_4R$, Sigma1, $\rm D_3R$, and GRM2, based on the corresponding predicted BA values of -9.61 kcal/mol, -9.55 kcal/mol, -9.55 kcal/mol, and -9.37 kcal/mol. In addition, the BA value to \textit{hERG} is predicted to be -7.80 kcal/mol.

Atomoxetine is a selective NET inhibitor and has been approved for the treatment of ADHD, and recently found to prevent relapse to cocaine use. Other preclinical studies have shown that atomoxetine can significantly attenuate cue-induced relapse to cocaine seeking after abstinence, which reflects the potential of atomoxetine as an effective treatment in preventing relapse in cocaine addiction \cite{economidou2011selective}. Safety studies were carried out on atomoxetine when used with intravenous cocaine on cocaine-experienced participants and found that atomoxetine can be safely tolerated \cite{cantilena2012safety}. Using our ML models, it  is predicted to have BAs of -9.27 kcal/mol, -9.21 kcal/mol, and -9.33 kcal/mol respectively for SPR, APP, and LYN. Side effects to these targets, discussed above, may be anticipated. Atomoxetine is already an FDA-approved medication with low \textit{hERG} side effects.

\subsection{Nearly optimal leads from our systematic screening and repurposing}

We are dedicated to finding promising lead compounds targeting SERT and NET. Screening and repurposing are two key processes for us to filter molecules from 61 available inhibitor datasets, as summarized in Figure \ref{fig:cross-prediction-heatmap}. In addition to drug potency and side effects, we also consider the six ADMET properties in table \ref{tab:property-optimal} as well as synthetic accessibility. In total, we required 68 criteria to be satisfied for a drug to be considered as a potential reliable agent. In screening, we started with potent inhibitors (experimental BA values $<$ -9.54 kcal/mol) from the inhibitor dataset of the target protein. Then those potent inhibitors were examined for their side effects on the other 59 proteins with a uniform BA requirement, i.e. predicted BA values higher than -9.54 kcal/mol. A stricter BA threshold is applied for \textit{hERG}, with predicted BA values higher than -8.18 kcal/mol required. These predicted side effect BA values were obtained from our proteme-models of the 61 proteins. For repurposing, we started with inhibitors of low BA value (experimental BA $>$ -9.54 kcal/mol) to their designated target protein yet having high predicted BA ($<$ -9.54 kcal/mol) to SERT or NET. Following this, side effect examinations were carried out on the other 59 proteins. Finally, excellent ADMET properties and synthesizability specified in table \ref{tab:property-optimal} had to be satisfied for both processes, -9.54 kcal/mol) to SERT or NET. Following this, side effect examinations were carried out on the other 59 proteins. Finally, excellent ADMET properties and synthesizability specified in table 1 had to be satisfied for both processes.

Several compounds through screening or repurposing were found as potential agents for SERT. They all have excellent ADMET properties and are readily prepared. Compound CHEMBL1411979 was obtained via screening from the SERT inhibitor dataset. It has experimental BA of -9.54 kcal/mol to SERT, and its potential side effects on \textit{hERG} are low with a predicted BA of -7.15 kcal/mol. Moreover, its predicted BA is stronger than cocaine (experimental BA value -9.08 kcal/mol). It deserves further investigation. 7 additional compounds were predicted to have low side effects against the other 59 targets. Among those, CHEMBL3800268, from the LRRK2 inhibitor dataset, has a low BA of -8.35 kcal/mol to its designated target, but it is predicted to have BA of -10.01 kcal/mol to SERT. Its potential side effects on other proteins are weak. For example, its predicted BA to \textit{hERG} is as low as -7.12 kcal/mol. Another compound, CHEMBL270299, from $\rm D_3R $ dataset is predicted to have a BA of -9.62 kcal/mol for SERT. It has a weak potential \textit{hERG} side effect with a predicted BA -7.57 kcal/mol, while it has a low experimental BA of -8.08 kcal/mol to its designated target $\rm D_3R $. The information on other compounds and their predicted BAs can be found in the supporting information. All these compounds are predicted to have more potent BAs to SERT than that of cocaine.

In searching for effective inhibitors for NET, compound CHEMBL454675 was found to satisfy all requirements. It has an experimental BA of -11 kcal/mol for NET while its predicted BA to \textit{hERG} is just -7.52 kcal/mol. It may provide good pharmacological effects, as its BA for NET is much higher than that of cocaine (-9.33 kcal/mol). No compounds satisfying our criteria were found for NET by repurposing from the other 60 inhibitor datasets.

Figure \ref{fig:docking} provides the docking information of cocaine, Ibogaine, and two nearly optimal lead compounds at the central site of SERT. The residuals at active sites near these compounds are also labeled. More docking information about other nearly optimal lead compounds can be found in the supporting information. These docking predictions were implemented by software AutoDock Vina \cite{trott2010autodock}. 

\section{Conclusion}

Substance use disorder (SUD) is associated with a variety of mental/emotional, physical, and behavioral problems, including chronic guilt, the seeking and taking of drugs despite adverse consequences, driving or making important decisions while intoxicated, and physiological withdrawal symptoms. As a specific example of SUD, millions of people are addicted to cocaine. However, there isn’t currently a therapeutic approved by the U.S. Food and Drug Administration to address, in part because cocaine addiction involves intricate molecular mechanisms. DAT, SERT, and NET are each associated with a complex interactome network, and one cannot develop anti-cocaine addiction medications without taking into account all interactome networks.

We have proposed proteome-informed machine learning studies of cocaine addiction as the first interactome-based machine learning/deep learning (ML/DL) protocol for anti-cocaine addiction lead discovery \cite{gao2021proteome}, but only the DAT interactome network was considered in this work. The present work extends these previous studies to SERT and NET, enabling us to perform a comprehensive evaluation of existing potential cocaine addiction inhibitors. We have considered repurposing existing inhibitors and screened for side effects as well as ADMET properties. After this rigorous screening, we have identified a small group of promising compounds. 

The knowledge and understanding obtained from the present work will be employed for the automated generation and screening of anticocaine addiction candidates using our generative network complex \cite{gao2020generative}. The next step is to test the resulting leads in \textit{in vitro} and animal assays. It will be critical to examine toxicity and blood-brain barrier permeability characteristics of candidate compounds using cell-based assays to refine our lists and prioritize compounds for animal models. This may include iterative medicinal efforts to optimize  critical qualities to identify compounds with the greatest therapeutic potential. These compounds must then be tested in rodent models, as \textit{in silico} and cell-based assays cannot definitively determine the behavioral effects of a drug. Cocaine locomotor sensitization in mice can uncover the ability of a compound to block the physiological and psychomotor effects of the drug while cocaine conditioned place preference can determine whether a compound can prevent drug reward or drug-environment \cite{gajewski2019epigenetic,doyle2021serum,prus2009conditioned}. Compounds with effects in either of these assays would then be good candidates for testing in the more time-consuming but more rigorous cocaine self-administration model. Examining characteristics like acquisition, breakpoint, extinction, and context- or cue-reinstatement can reveal whether a compound might be useful to block key aspects of cocaine addiction like drug seeking, craving, and relapse \cite{doyle2021serum,thomsen2007intravenous}. Of course, compounds that produce promising results in animal testing would then move on to clinical trials.

Finally, our work establishes a new protocol for artificial intelligence (AI)-based nearly optimal lead discovery that can be applied to any disease for which some portion of the molecular etiology has been studied. We hope this technology can be applied to many other neuropsychiatric diseases going forward to uncover new classes of therapeutic agents to improve disease outcomes.

\section*{Data  and model availability}

 The 61 cocaine-addiction related datasets studied in this work are publicly available at:\\
https://weilab.math.msu.edu/DataLibrary/2D/. \\
Our source code and trained autoencoder model for LV-FP generation can be found at \\
https://github.com/WeilabMSU/antoencoder-v01.

\section*{Acknowledgment}
This work was supported in part by NIH grants GM126189 and DA040621, NSF Grants DMS-2052983, DMS-1761320, and IIS-1900473, NASA 80NSSC21M0023, MSU Foundation, Michigan Economic Development Corporation, George Mason University award PD45722, Bristol-Myers Squibb 65109, and Pfizer.



\end{document}